\pdfoutput=1
\documentclass[amsart, aps,prd,groupedaddress]{revtex4}
\usepackage{graphicx,subfigure,enumerate,pstricks,latexsym,longtable}
\usepackage[english]{babel}
\usepackage{epstopdf}
\usepackage{hyperref}
\usepackage{amsmath}
\usepackage{amsfonts}
\usepackage{systeme}
\usepackage{leftidx}
\usepackage{xcolor}

\newcommand*{\Scale}[2][4]{\scalebox{#1}{$#2$}} 

\begin{document}

\title{A topological approach for emerging D-branes and its implications for gravity }
\author{Richard Pincak$^{1,2}$ \quad Alexander Pigazzini$^{3}$ \quad Saeid Jafari$^{3,4}$ \quad Cenap {\"{O}}zel$^{5}$ \quad Andrew DeBenedictis$^{6}$ 
\\
$^1$ {\it Institute of Experimental Physics, Slovak Academy of Sciences, Kosice, Slovak Republic; pincak@saske.sk}\\
$^2$ {\it Joint Institute for Nuclear Research, Dubna, Russia}\\
$^3${\it Mathematical and Physical Science Foundation, 4200 Slagelse, Denmark; pigazzini@topositus.com}
$^4${\it College of Vestsjaelland South, Herrestraede 11, 4200 Slagelse, Denmark; saeidjafari@topositus.com}\\
$^5${\it King Abdulaziz University, Department of Mathematics, 21589 Jeddah, Saudi Arabia; cozel@kau.edu.sa}\\
$^6${\it The Pacific Institute for the Mathematical Sciences, and Department of Physics, Simon Fraser University, V5A 1S6 Burnaby - B.C., Canada; adebened@sfu.ca}}

\begin{abstract}
We introduce  a new geometric/topological approach to  the emerging braneworld scenario  in the context of D-branes using partially negative dimensional product (PNDP) manifolds. The working  hypothesis  is based on the fact that the orientability of PNDP manifolds can be arbitrary, and starting from  this,  we propose that gravitational interaction can derive naturally from the non-orientability. According to this hypothesis, we show that topological defects can emerge from non-orientability and they can be identified as gravitational interaction at  macroscopic level. In other words,  the orientability  of  fundamental PNDPs  can be related to the appearance of curvature at low energy scales. 
\\
\\
\textup{Mathematics Subject Classification 2020: 83E30; 53C25} 
\\
\textup{PACS numbers: 11.25.Yb; 11.25.-w; 02.40.Ky; 02.40.-k; 04.50.-h} 
\\
\textup{Keywords: Braneworld, dimensionality, topology, orientability, gravity.}
\end{abstract}

\maketitle

\section {Introduction and Preliminaries}

\begin{figure}[h!t]
	\centering
	\includegraphics[width=0.85\textwidth]{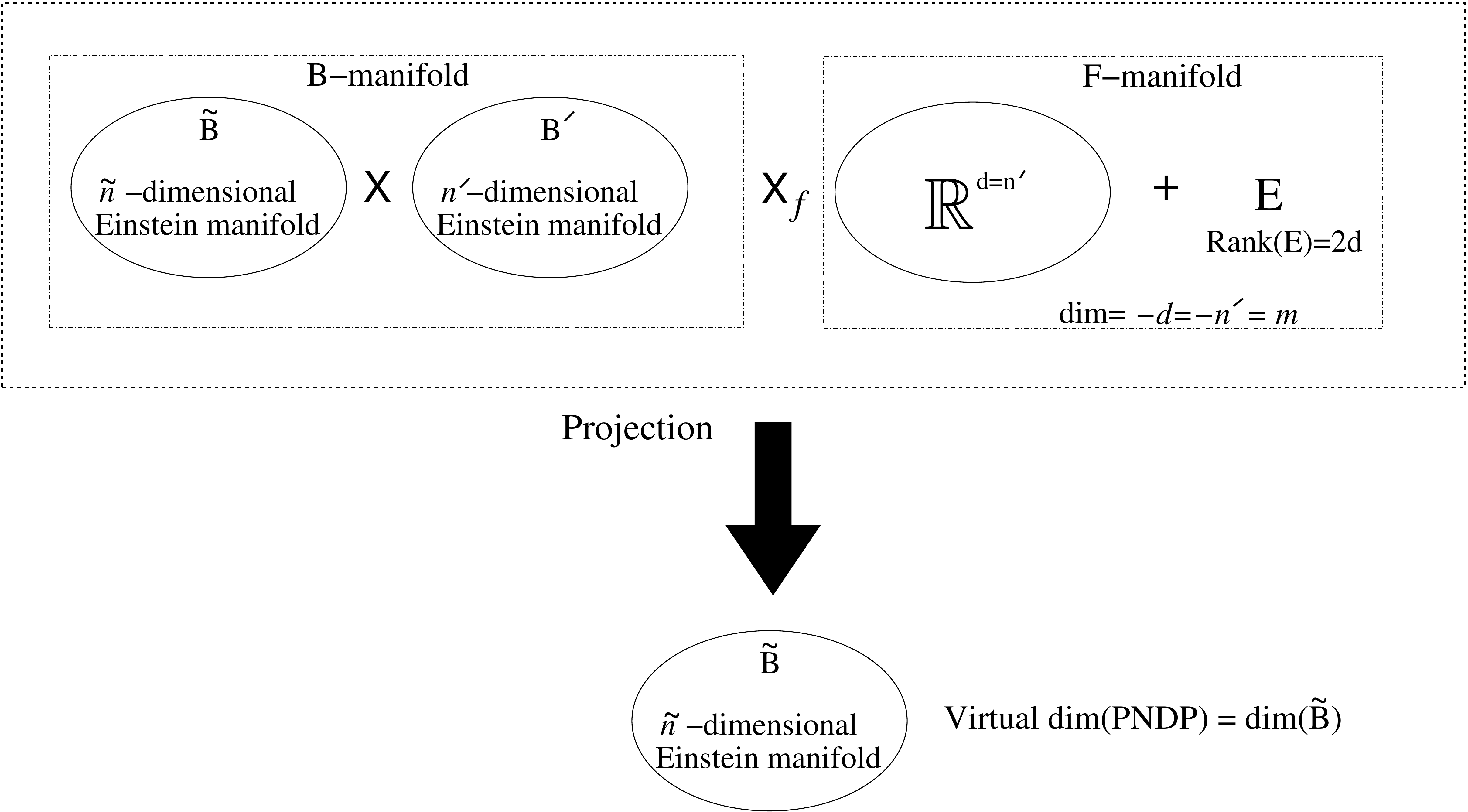}
	\caption{\small{Graphical representation of a PNDP-manifold with an interpretation of ``virtual" dimensions.}}
	\label{fig:PNDP_manifold_1}
\end{figure}

In string theory, D-branes, short for Dirichlet membrane, are a class of extended objects upon which open strings can end with Dirichlet boundary conditions, after which they are named. D-branes were discovered by Dai, Leigh and Polchinski (see \cite{Dai}), and independently by Horava \cite{Horava} in 1989.
\\
The study of D-branes (see \cite{Johnson}) introduced the concept of the brane world and thus the brane world cosmology. Because string theory implies that the Universe has more dimensions than we expect, we have to find a reason why the extra dimensions are not apparent. One possibility would be that the visible Universe is in fact a very large D-brane extending over three spatial dimensions. Material objects, made of open strings, are bound to the D-brane, and cannot move off of the brane to explore the Universe outside the brane. This scenario is called a brane cosmology. The force of gravity is not due to open strings; the gravitons which carry gravitational forces are vibrational states of closed strings. Because closed strings do not have to be attached to D-branes the gravitational effects could depend upon the extra dimensions orthogonal to the brane. According to this picture, effective theories of gravity can emerge giving rise to observational effects \cite{Basini}, \cite{Basini2}, \cite{Report}.
\\
In this paper we intend to introduce a new way to understand the braneworld using the theory of emergent spaces (or PNDP-Theory, for ``partially negative dimensional product'' manifold \cite{pndp}, \cite{pndp-string}) and specifically what we call an ``emerging braneworld'' using a topological approach. For doing this we have used PNDP-manifolds.
In addition to the mathematical aspect, which concerns the construction and studying of these new types of Einstein warped products manifolds, our physical interpretation of the interactions between positive dimensions and ``virtual'' negative dimensions lends these manifolds to a variety of applications. In this paper we deal with some explicit constructions of branes.
\\
To present these applications, we take as basic knowledge that which is stated in \cite{pndp} and \cite{pndp-string}. In the theory presented there it is proposed that the fundamental structures in the universe are dimensions, and everything can in principle be built out of these dimensions via topological considerations. Even the strings themselves arise from dimensions. 
\\
Let us recall briefly, from \cite{pndp}, that a PNDP-manifold is a manifold $(M, \bar{g})=(B, g) \times_f (F, \ddot{g})$, where the base-manifold $(B, g)$ is a Riemannian (or pseudo-Riemannian) product-manifold, which we can write as $B=B'\times \widetilde B$, with $g=\Sigma g_i$, where $\widetilde B$ is an Einstein manifold, and the fiber-manifold $(F, \ddot{g})$ is a derived differential Riemann-flat manifold with negative ``virtual'' integer dimensions $m$, where with \textit{derived differential manifold} is considered  a \textit{smooth Riemannian flat manifolds by adding a vector bundle of obstructions, $\mathbb{R}^d+E$}. In particular we  consider, for $F$, only the spaceforms $\mathbb{R}^d$, with orthogonal Cartesian coordinates such that $g_{ij}=-\delta_{ij}$, by adding a vector bundle of obstructions, $E \rightarrow \mathbb{R}^d$, (so $F:=\mathbb{R}^d + E$, where $dim(F)=m=d - rank(E)$, such that $m=-d$). In fact in this circumstance, if we consider a Kuranishi neighborhood $(\mathbb{R}^d, E, s)$, with manifold $\mathbb{R}^d$, obstruction bundle $E \rightarrow \mathbb{R}^d$, and section $s: \mathbb{R}^d \rightarrow E$, then the dimension of the derived manifold $F$ is $dim(\mathbb{R}^d) - rank(E)$. For more information about Kuranishi neighborhoods (derived smooth manifolds and obstruction bundle) see \cite{Joyce}. Schematically the situation is illustrated in figure \ref{fig:PNDP_manifold_1}.

That said, let's also remember that  a PNDP-manifold satisfies the following system (see \cite{pndp}, \cite{pndp-string}):
\\
\\
\begin{equation}
\bar{Ric}=\lambda \bar{g} \Longleftrightarrow\begin{cases} 
 Ric'- \frac{d}{f}\tau'^*\nabla'^2 f'= \lambda g' \\ \widetilde \tau^* \widetilde \nabla^2 \widetilde f=0
\label{eq:conditions1}\\
\widetilde Ric = \lambda \widetilde g \\ \ddot{Ric}=0 \\ f \Delta' f'+(d-1) |\nabla f|^2 + \lambda f^2 =0,
\end{cases}
\end{equation}
\\
\\
(since $Ric$ is the Ricci curvature of $B=B' \times \widetilde B$, then $Ric=Ric' \oplus \widetilde Ric$, so $Ric=\lambda(g'+\widetilde g)+ \frac{d}{f'}\tau'^*\nabla'^2 f'$),
with $Ric'$ is the Ricci tensor of $B'$, $\widetilde Ric$ is the Ricci tensor of $\widetilde B$, $\bar{Ric}$ is the Ricci tensor of the whole PNDP-manifold, $\ddot{Ric}$ is the Ricci tensor of $F$, $g'$ is the metric tensor referred to $B'$. The quantity $\widetilde g$ is the metric tensor referred to $\widetilde B$, $\bar{g}$ is the total metric, $f(x,y)=f'(x)+ \widetilde f(y)$  is the smooth warping function $f:B \rightarrow \mathbb{R}^+$ (where each is a function on its individual manifold, i.e., $f':B'\rightarrow \mathbb{R}^+$, and $\widetilde f:\widetilde B \rightarrow \mathbb{R}^+$), $\nabla^2 f=\tau'^*\nabla'^2f+\widetilde \tau^* \widetilde \nabla^2 \widetilde f$ is the Hessian referred on its individual metric, (where $\tau'^*$ and $\widetilde \tau^*$ are the pullbacks), $\nabla f$ is the gradient (then $|\nabla f|^2= |\nabla' f'|^2 + |\widetilde \nabla \widetilde f|^2$), and $\Delta f=\Delta' f'+ \widetilde \Delta \widetilde f$ is the Laplacian, (since $\widetilde B$ is Einstein, we have $\widetilde \tau^* \widetilde \nabla^2 \widetilde f=0$, then $\widetilde \Delta \widetilde f=0$).

Therefore, contracting the first equation of (\ref{eq:conditions1}), for $dim(M)=0$ and $dim(M)<0$ (with $M$ we mean the whole PNDP-manifold) , the system equations becomes:
\begin{equation}
\bar{R}=\lambda \bar n \Longleftrightarrow\begin{cases} 
R'f- \Delta' f'd=n' f \lambda \\ \widetilde \Delta \widetilde f=0 \\ \widetilde R = \lambda \widetilde n \\ \ddot{Ric}=0 \\ f \Delta' f'+(d-1) |\nabla f|^2 + \lambda f^2 =0, \label{eq:conditions2}
\end{cases}
\end{equation}
\\
\\
where $n'$ and $R'$ are the dimension and the scalar curvature of $B'$ respectively, and $\widetilde n$ is the dimension of $\widetilde B$; while for $dim(M)>0$, as per definition of PNDP-manifold (see \textit{Definition 4} in \cite{pndp}), we must set $d=n'$ (we remember that $d$ is the dimension of the underlying manifold ($R^d$) that defines the fiber-manifold $F$). We obtain for this case:
\\
\begin{equation}
\bar{R}=\lambda \bar n \Longleftrightarrow\begin{cases} 
R'f- \Delta' f'n'=n' f \lambda \\ \widetilde \Delta \widetilde f=0 \\ \widetilde R = \lambda \widetilde n \\ \ddot{Ric}=0 \\ f \Delta' f'+(n'-1) |\nabla f|^2 + \lambda f^2 =0. \label{eq:conditions3}
\end{cases}
\end{equation}
\\
\\
Finally with regard to the projections/desuspensions we remember the following points:
\\
- if $dim(M)>0$ (i.e. system solutions (\ref{eq:conditions3})) we have the projection:
\\  
\begin{equation}
\pi_{(>0)}:PNDP\rightarrow (\Pi_{i=(q'+1)}^{\widetilde q}B_i)=\widetilde B,
\end{equation}
\\
- if $dim(M)=0$, (i.e., system solutions (\ref{eq:conditions2})), we have the projection:
\begin{equation} 
\pi_{(=0)}:PNDP\rightarrow P, 
\end{equation}
where with $P$ we mean a point-like manifold, of zero dimension, and
\\
- if $dim(M)<0$, (i.e., system solutions (\ref{eq:conditions2})), we have the projection:
\begin{equation}
\pi_{(<0)}:PNDP\rightarrow \Sigma^{dim(M)<0}(p), 
\end{equation}
with $\Sigma^{dim(M)<0}(p)$, we mean the $(||dim(M)||)$-th desuspension of point, for example, if $dim(M)=-4$ the projection $\pi_{-4}$ will project $M$ into an object which will be given by the fourth desuspension of a point.
\\
\\
As seen in \cite{pndp-string}, in the authors' approach, we consider the PNDP-manifolds as strings whose emergent parts are topologically equivalents to open or closed strings.
\\
For instance, we consider a PNDP-manifold with $dim(B) = 2$, where $B =(I_1 \times I_2)$, with $I_i \subset \mathbb{R}$, or $B =(S^1_1 \times S^1_2)$, where $S^1_i$ are circles. The fiber-manifold is $F=(\mathbb{R}+ E)$ with $rank(E)=2$, so $dim(F) = -1$. We obtain $dim(PNDP) = 2 -1 = 1$, i.e. our PNDP-manifold will emerge as a one-dimensional {``}object{''} topologically equivalent to an open or a closed string; in other words, considering the projection seen above, $\pi_{(=1)}:[(I_1 \times I_2)\times (\mathbb{R}+E)] \rightarrow I_1$, or $\pi_{(=1)}:[(S^1_1 \times S^1_2)\times (\mathbb{R}+E)] \rightarrow S^1_1$, respectively.
\\
\\
Therefore, in this approach, to obtain the dimensional equivalent of a string, we need $3$-dimensional space, in which one of these dimensions should be virtual negative, and only $1$-dimension of these emerges as an open or closed string. 
\\
So, the string is $(2-1)$-PNDP $=(I_1 \times I_2)\times F$ (or $(S^1_1 \times S^1_2)\times F$), and, for our interpretation, the emergent space, by our projection, will be $I_1$, i.e., a line interval, topologically equivalent to an open string, or $S^1_1$, i.e., a circle, topologically equivalent to a closed string.
\\
Since the PNDP-manifold-string is a Riemannian product of 1-dimensional manifolds, i.e., $(2-1)$-PNDP $=(I_1 \times I_2)\times F$ (or $(S^1_1 \times S^1_2)\times F$), and since the Ricci curvature of a 1-dimensional manifold is zero ($Ric'= \widetilde{Ric}=\ddot{Ric}=0$), its Ricci curvature will be $\bar{Ric}=0$ (and from $\bar{Ric}=\widetilde{Ric}=0$, then also $\lambda=0$), so the system (\ref{eq:conditions3}) becomes trivial:
\begin{equation}
\bar{R}=0 \Longleftrightarrow\begin{cases} 
\Delta' f'=0\\ \widetilde \Delta \widetilde f=0 \\ \widetilde R = 0 \\ \ddot{Ric}=0 \\ R'=0, \label{eq:zeroricconditions}
\end{cases}
\end{equation}
and we have that $\widetilde f$ and $f'$ (which are both constant such that $f=\widetilde f+f'=1$) are solutions for (\ref{eq:zeroricconditions}).
\\
In particular, a string is generated by the interaction that occurs between the dimensions of $(2-1)$-PNDP-manifold, that is two positive dimensions and a virtual negative dimension.
\\
To operate in the context of strings, we replace $(\mathbb{R}+ E)$ with $(I_3 + E)$ where $I_3$ is a linear interval of $\mathbb{R}$ with string length, so the two types of interactions will therefore be the following:  
\\
\\
$Int_A= I_2\times (I_3+E)$, and
\\
$Int_B= S^1_2\times (I_3+E)$,
\\
\\
where $Int_A$ is the interaction present in $(I_1 \times I_2) \times (I_3 +E)$, i.e. which determines the open string, while $Int_B$ is the interaction present in $(S^1_1 \times S^1_2)\times (I_3+E)$, i.e. which determines the closed string.

\section{PNDP-D-branes}

We consider the following hypothesis as a possible scenario.
\\
\\
\textit{Hypothesis}: Gravitation could derive directly from the fundamental PNDPs. In fact within the PNDP the interacting dimensions emerge as point-like manifolds, that is as points. Assuming that these can join together, they would compose a space with topological defects that could mimic gravity at a fundamental level. Gravity, fundamentally in this hypothesis, is due to topological effects present due to the PNDPs making up the branes.
\\
\\
\subsection{Development of the Hypothesis}

\begin{figure}[h!]
 \centering
  \includegraphics[width=0.3\textwidth]{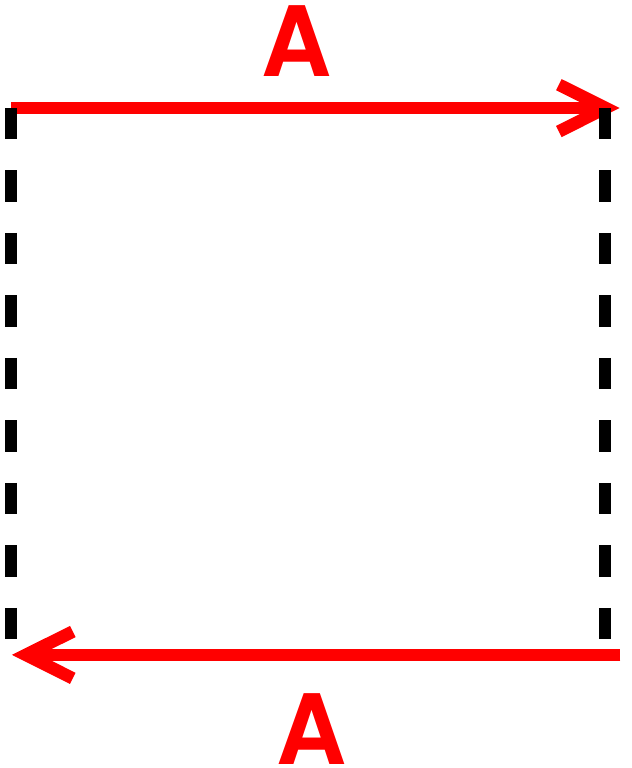}
\caption{\small{To turn a square into a M\"{o}bius strip, flip one of the edges labeled A so that the arrows point in the same direction.}}
\label{picture_square_2}
 \end{figure}

\begin{figure}[h!t]
\centering
\includegraphics[width=1.0\textwidth]{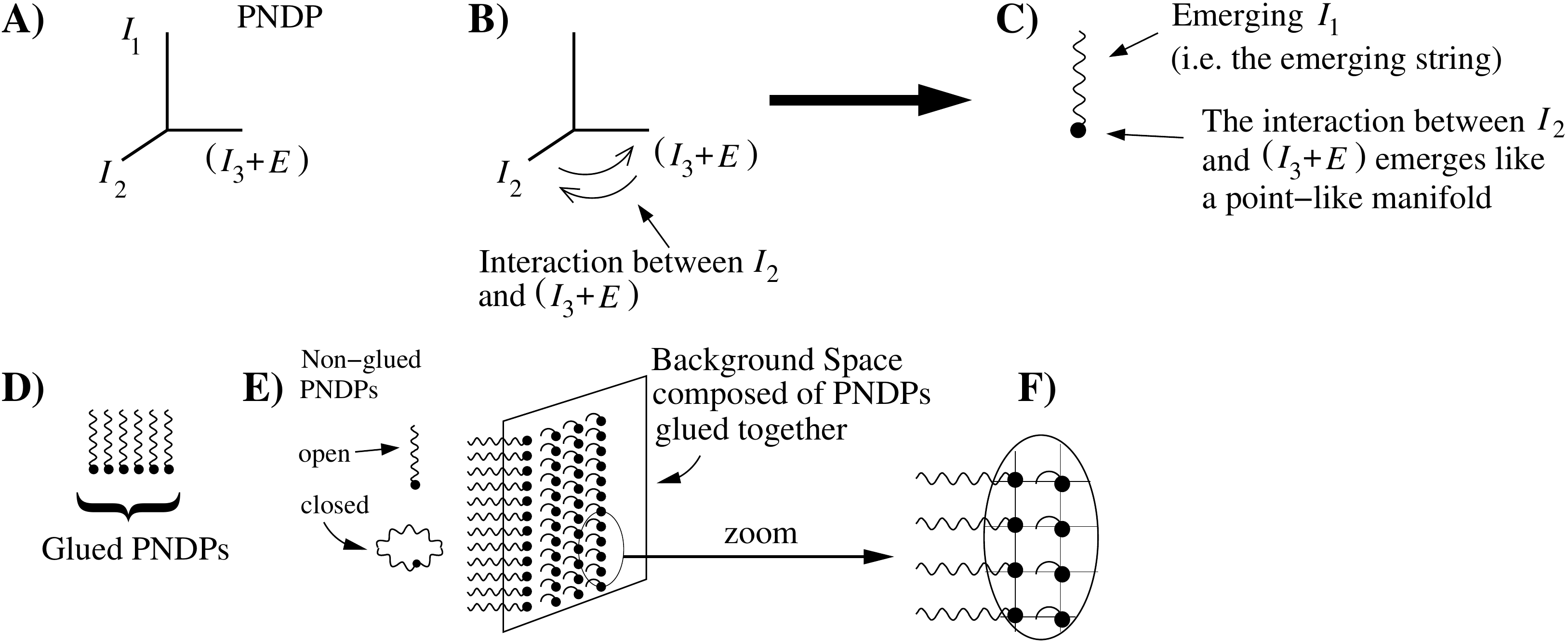}
\caption{\small{The construction of a brane via PNDPs.}}
\label{fig:glued_PNDPs_3}
\end{figure}

In PNDPs {``}open strings", $I_1 \times I_2 \times (I_3 + E)$, we suppose that $I_2 \times (I_3 + E)$ has the ``upper" side identified with the ``lower" side, we would have that the latter would generate a sort of tape that emerges point-like with a hooked string. In other words, the two dimensions interact via the theory of PNDPs and in this way the individual PNDP manifolds, which are perceived as point-like, actually possess internal properties, such as orientability, of a two dimensional manifold.
\\
If in the $I_2 \times I_3$ manifold, the sides are identified by the relation $(x,0)\sim (\epsilon-x,\epsilon)$ for $0\leq x\leq \epsilon$ (where by $\epsilon$ we mean the string length), it will be a M{\"{o}}bius strip, while if the sides are identified with the relationship $(x, 0) \sim (x, \epsilon)$, so without  a half turn of twist, then we have an oriented strip, as with a cylinder. We show this schematically in figure \ref{picture_square_2}.
\\
\\
Therefore, in our hypothesis, we exploit the fact that the interacting dimensions of the PNDP-manifolds can originally be configured in such a way as to have an arbitrary orientability compatible with two dimensional manifolds. So, we will examine the two simplest cases: $(I_2 \times I_3)$ are oriented manifolds (oriented strips, i.e., cylinders), and $(I_2 \times I_3)$ are non-oriented manifolds (i.e., M{\"{o}}bius strips). So we have that the interacting manifolds (i.e., $I_2 \times I_3$) can be M{\"{o}}bius strips (non-orientable), or else we can also have oriented strips (i.e., cylinders), and we will have that they emerge as point-like manifolds (because we have $I_2 \times (I_3 + E)$, which has virtual dimension equal to zero). We will call these either ``non-oriented point-like manifolds" or ``oriented point-like manifolds" depending on whether they are the emergent result of a M{\"{o}}bius strip, or of an oriented strip, respectively. 
\\ 
Now, following our hypothesis, we will have that our D-brane will be composed of {``}oriented point-like" and {``}non-oriented point-like manifolds" glued together (each of which has a string attached, i.e. $I_1$, the one-dimensional manifold that emerges from a PNDP), and we want to consider that the non-orientability of the point-like manifold corresponds to identifying the PNDP as a topological defect of the D-brane. 
\\
\\
Continuing to follow this hypothesis, we can try to adapt the concept of an action to our emerging approach. For doing this we must consider that our emergent string is actually a part of a three-dimensional structure in which the two interacting manifolds (we are referring to the $1$-dimensional manifold $I_2$ and to the $1$-dimensional manifold $(I_3+E)$), which are part of the structure, are not perceived because, as already mentioned, they cancel each other out giving rise to a point-like manifold via PNDP theory. The string, however continues to be attached to them. We hypothesize that these structures (PNDPs) have the possibility of joining together to form more complex structures, much like a line may be perceived to be constructed out of points. In fact, since PNDPs emerge as a string attached to a point-like manifold (the point-like manifold is how the interaction emerges), they can be glued together (and even unglued) along the point-like manifolds, forming a D-brane. We therefore consider the D-brane (M) (composed of the union of the PNDPs as hypothesized above), which has a flat fundamental connection $\omega$. By flat fundamental connection we mean that the connection at the microscopic level is must be flat due to the $R^{1}\times R^{1}$ structure of the part of the PNDPs making up the brane. Clearly the conformal symmetries of related fields have to be considered in the construction \cite{Casalbuoni}.
\\
\\
As previously said, we consider topological defects, that is, the presence of non-orientable PNDPs, and for simplicity we will refer to them as PNDPs whose interacting manifolds (i.e., $I_2 \times (I_3 + E)$) are M{\"{o}}bius strips, but in reality they could be two-dimensional Klein bottles, projective planes, etc., that is, non-orientable surfaces that admit a flat connection.
\\
\\
In figure \ref{fig:glued_PNDPs_3}, especially in (F), the structure of the D-brane is shown. All physical space is created out of the gluing of PNDP-manifolds, i.e., the D-brane, which emerges dense out of point-like manifolds.

\bigskip
In this approach, the concept of gravity is introduced by the presence of ``M\"{o}bius defects'', specifically the presence of non-orientable point-like manifolds is considered as an indication of the presence of gravity at the microscopic level. Therefore gravity is considered a topological defect in the ``fundamental orientability" (see also \cite{Guo} and \cite{Hadley}).
\\
So, the more M{\"{o}bius-PNDPs are present in a portion of the D-brane, the more we will say that gravity is present in that portion of the D-brane. This will be elaborated on below in the discussion of holonomy at the macroscopic level.
\\
\\
Based on our hypothesis, we consider that a D-brane is composed of only ``open PNDP-strings" (with open PNDP-string we mean the usual PNDP manifold that emerges as point-like manifold with an attached open string). There can also be other open PNDP-strings and closed PNDP-strings moving freely, as in figure \ref{fig:glued_PNDPs_3}E.
\\
\\
{\bfseries NOTE:} \textit{To recap, the interaction we consider that emerges point-like, has dimension given by $1+ (-1)=0$, and regarding $F$, it must be said that in this case it has dimension $1 + (- 2) = -1$, where ($-2$) is the rank of the bundle of obstruction ($rank(E)=-2$), so $dimF = -1$.
\\
Therefore we have the interaction: $I_2 \times F$, that is $I_2 \times (I_3 + E)$ which has zero dimension: $1 + (- 1) = 0$, but if we consider only the underlying manifold of $F$, (that is $I_3$) it has dimension $1$, so if we consider $I_2 \times I_3$ then we have dimension $1 + 1 = 2$.
\\
Considering only the Riemannian product with the underlying manifold of $F$ and the manifold $I_2$, we will have the M{\"{o}}bius strip, or else we will have an oriented strip. Considering the algebra on $F$ (the bundle of obstruction) we have $I_2 \times (I_3 + E)$, i.e., its dimension is $1 + (- 1) = 0$, point-like manifold. 
Then we say that the point-like portion also contains the information of the underlying structure, because it is the same object, and for an observer the point-like manifold is only how it emerges}.
\\
\\
{\bfseries Remarks 1:} \textit{Being that PNDP-manifolds can glue and unglue to create D-branes (see figure \ref{fig:picture_4}), there is no reason to believe that there cannot be more D-branes and even an ambient space (Bulk) composed of PNDP-manifolds that contains a multitude of D-branes. For this reason we consider the D-brane under consideration as a submanifold of a $\mathbb{R}^n$ space (Bulk)}.

\begin{figure}[h!]
 \centering
  \includegraphics[width=0.6\textwidth]{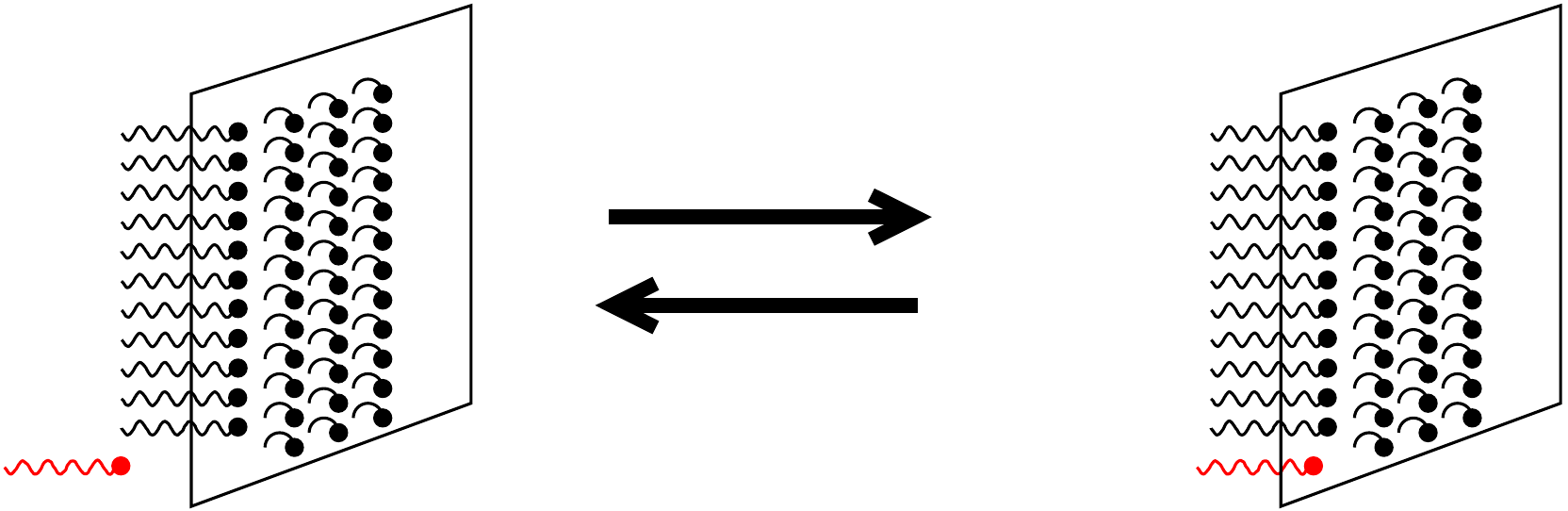}
\caption{\small{Glued and unglued point-like manifolds modify the portions of D-brane, creating dynamics.}}
\label{fig:picture_4}
 \end{figure}

\subsection{From gravity index  at microscopic level  to curvature  at macroscopic level}

The approach considered so far concerns the microscopic aspect, in which we consider the appearance of gravity as a topological defect (the presence of non-orientable point-like manifolds).
\\
At a certain point we believe it appropriate that the approach illustrated so far should be united with General Relativity and for this reason we must show that the non-orientability at the fundamental level of the point-like manifolds create an effective curvature at the macroscopic level, compatible, in the low energy scale, with General Relativity.
\\
\\
In this regard, we want to attribute a geometric role to the interactions that occur between the underlying $1$-dimensional manifolds of the emerging point-like manifold, thus introducing the possibility that it is precisely the interactions that determine a defect that we interpret as a ``parallelism defect " of a vector transported along a closed loop on the emergent D-brane.
\\
This contribution concerns the rescaling of the angular phase shift, obtained from the orientability (zero radians, therefore zero contribution), or non-orientability ($\pi$ radians) of the underlying surface of the point-like manifold, which will modify, point-like manifold after point-like manifold, the direction of the vector transported, obtaining a parallelism defect on our D-brane.
\\
Since we are in an unspecified codimension, but certainly greater than $1$, then for the normal vector to the underlying surface of the point-like manifold, we will not have a canonical choice, so we will mean any vector lying in the normal bundle, which can obviously be defined.
\\
\\
To begin, we consider that our D-brane is immersed in $\mathbb{R}^n$ (bulk), and let's say that the interactions of the underlying two-dimensional manifold of the emerging point-like manifolds, on a large scale (macroscopic level) mimic an effective connection (although the fundamental connection, $\omega$, is flat), and to do this we consider the following steps:
\\
\\
(1) We have a flat connection $\omega$ at the microscopic level.
\\
(2) The interacting dimensions of the PNDPs can either possess a  M{\"{o}}bius orientability, or else could be an oriented band.
\\
\\
(3) The M{\"{o}}bius case introduces a topological defect and this is identified with gravity on the large scale.
\\
\\
(4) We transport a vector along a small closed loop on the D-brane.
\\
To describe what we interpret as a parallelism defect, we must consider (as evidenced in \textit{NOTE} above), that the point-like manifold is actually just one way in which the $I_2 \times I_3$ manifold emerges (which can be M{\"{o}}bius strip or oriented strip).
Therefore a vector transported along a closed curve on the D-brane, for each point-like manifold (which makes up this curve) will undergo the influence of the interaction (which occurs between the two 1-dimensional manifolds that make up the underlying surface of the point- like manifold). This acts such as to modify its direction through the contribution given by the phase shift (caused by the non-orientability of the underlying surface). 
\\
\\
(5) Returning to the interactions, we want to associate to them a geometric contribution, that is $\pm \mu \beta$, equal to the rescaling of the phase shift of the angle obtained from the non-orientability of the underlying surface of the point-like manifold. That is, considering, as already said, that the point-like manifold and this underlying surface are the same thing, then we imagine that a vector placed on the point-like manifold in question can experience its orientability and to do this we consider that it {``}falls" in the normal direction on the underlying surface. If this surface is non-orientable (M{\"{o}}bius strip), then, as is well known, starting from a point and continuing along the entire strip, the normal unit vector arrives at the starting point with an inverted orientation, i.e. phase shift of $\pi$ radians, as can also be seen from the holonomy group of the M\"{o}bius strip (i.e., $\mathbb{Z}/2\mathbb{Z}$). The rescaling considered causes this angle to become: $\alpha = \pm \mu \beta$ (where $\mu$ is a very small value and is the only constant that needs to be specified at this stage for the theory). $\beta$ is the phase shift of the angle obtained (that is $0$ radians (oriented strip) or $\pi$  radians (M{\"{o}}bius strip)) and this helps to determine the direction that the vector assumes during the transport that we perform on the D-brane. 
\\ 
Therefore, considering what has been said, if the underlying surface of the point-like manifold were a M{\"{o}}bius strip, then the contribution of phase shift, due to the interactions (which will help to determine the direction of the vector in that point-like manifold), will be equal to $\pm \mu \pi$.
\\
Before continuing, we must make a premise, namely that we do not know a priori the effective connection at the macroscopic level, but we know that at the microscopic level $\omega$ is flat. Therefore being that the parallelism defect and the consequent connection are only effects that are mimicked by the D-brane, we consider that in reality everything is flat and only at a macroscopic level will we have the illusion of the presence of an effective curvature. In this regard we want to consider that even a curve placed on the D-brane is a plane curve, that is, trivially contained in a 2-dimensional plane. Only afterwards at a macroscopic level will the effect be perceived that will mimic the non-planarity of the closed loop. This could mean that the D-brane is actually $\mathbb{R}^D$ and that only at the macroscopic level will it give us the illusion of being a curved manifold. 
\\
Having said this, to determine the direction of the vector on the first point-like manifold, and consequently also to determine the entire parallel transport, we consider the following steps: 
\\
\\
- We consider the closed plane curve and a tangent vector $w$ to it in the point-like manifold. This vector will belong to the same plane that contains the closed plane curve. We modify the direction of this vector by adding to it the rescaled contribution ($\pm \mu \beta$), due to the interactions on the underlying surface, so that the vector obtained, which we call $w'$, will no longer be tangent to the curve, but will still belong to the plane containing the closed curve.
\\
- Now we consider the vector $v$, given by the normal vector $n$ to the underlying surface of the point-like manifold to which we add the quantity $\pm \mu \beta$, so that $v$ is contained in the 2-plane $V$ formed by the span of the two vectors $w'$ and $n$.
\\
 - Therefore, the direction of the vector (which we call $w''$) on the first point-like manifold, which constitutes the closed curve on the D-brane, will be such that its scalar product with the vector $v$, just obtained, must be zero and at the the same time must remain contained in the $V$ plane. So $w''$ will be contained in $V$, but also in the 2-plane $W$, formed by all vectors whose scalar product with the vector $v$ is zero.  
\\
\\
So, the vector on the second point-like manifold will be such as to obey the following conditions:
\\
We consider the vector $w''$ on the first point-like manifold, therefore contained in the $W$ plane, and we add to it the new rescaled phase shift obtained from the underlying surface of the second point-like manifold, so that the new vector $w'''$ must be still contained in $W$. Let us now consider the vector $v$ and we also add to it the new rescaled phase shift from second point-like manifold, so that the obtained vector, called $v'$, belongs to the 2-plane $V'$, given by the span of the vectors $v$ and $w'''$. Then the vector $w''''$ on the second point-like manifold, will be such that its scalar product with $v'$ will be zero, but at the same time it must also be contained in $V'$ and in another 2-plane ($W'$), the latter formed by all vectors whose scalar product with $v'$ is zero. 
\\
In the same way we continue on the subsequent point-like manifolds until the closed loop is concluded, and this is what we consider parallel transport perceived at macroscopic level. 
\\ 
This phase shift, which we get between the initial vector and the one we get at the end of the small closed loop on the D-brane is a parallelism defect, and this mimics an effective curvature. In fact, as is known in differential geometry, the parallelism defect is attributable to the Riemann tensor, from which it is possible to go back to a connection (macroscopic level) and, assuming that there is a relationship between the connection and the metric, as in the Levi-Civita connection, we can in principle obtain a metric. In this way we show why the effective curvature is linked to gravity: by the presence of the non-orientability of what we call fundamental manifolds. In fact, having $\beta = 0$ for the oriented strips, we have a non-zero value only in the M{\"{o}}bius cases. Therefore, if completing the closed loop on the D-brane, when the vector encounters the point-like manifolds whose underlying surface is oriented strip it will not change its direction, while in cases where it meets the point-like manifolds whose underlying surface is M{\"{o}}bius strip, the vector will change its direction. 
\\
\\
(6) This means that the more M{\"{o}}bius strips we encounter along a closed loop on the D-brane, the more the vector will change direction, because a phase shift will be added going from point-like manifolds to point-like manifolds, making us obtain an effective curvature.  
\\
\\
(7) Thus, as mentioned, the important relationship between the fundamental orientability of PNDP manifolds and effective perceived curvature at the macroscopic level is obtained.
\\
\\
At the end of these steps, we can therefore summarize by saying that at the microscopic level the contribution given by the interactions that occur on the underlying surface of the point-like manifold is interpreted as a parallelism defect at a macroscopic level on the D-brane.
\\
Not knowing a priori the effective connection at the macroscopic level, we cannot parallel transport a vector. For this reason we want to consider the closed loop a plane curve and defining our way of parallel transporting (following the procedure described in point (5)), we can obtain a phase shift at the end of the closed loop.
\\
We interpret this phase shift at the end of the closed loop as a parallelism defect at the macroscopic level and from this defect we obtain an effective Riemann tensor, from which we can derive, in principle, the Christoffel symbols and therefore the connection at the macroscopic level. So, by ``calibrating" the connection on this phase shift, what we get is really an effective parallel transport of the vector. 
\\
So, for a ``citizen" of the D-brane (macroscopic level), the D-brane is perceived as a curved manifold.

\subsection{The Emerging Action}

At this point we describe the structure of an action suitable for the theory. To begin with, we can say that the goal of our action is to capture the dynamics of the D-brane portions.
\\
Being that the PNDP manifold, as mentioned, can glue and unglue thus modifying portions of the D-brane they create, a dynamic is induced in the portion of the D-brane. In practice, the Action measures the energetics of the hyper-volume of the portion of the D-brane that changes as the D-brane changes/deforms due fundamentally to the gluing and ungluing of PNDPs to create the volume of the brane.
\\
Having said that, let's consider a $X$ portion of D-brane contained in $R^n$ (Bulk), and its transformation as a result of glue/unglue, which we call $Y$.
We consider that there is a homotopic equivalence between $X$ and $Y$, i.e. there are two functions $f: X \rightarrow Y$ and $h: Y \rightarrow X$, such that $f \circ h$ is homotopic to the identity $Id_Y$ on $Y$ and $h \circ f$ is homotopic to the $Id_X$ identity on $X$.
We specify that we are not interested in describing this gluing and ungluing operation with homotopy equivalence at the microscopic level. We are only interested in showing that the portion of D-brane has dynamics, that is, from $X$ it passes to $Y$. In other words, the action here describes the macroscopic effects. In fact, by definition, a function between $X$ and $Y$ is continuous if the preimage of each open set of the interval space is open in the domain space. Even extending to a possible case where the glue/unglue operation renders one or both $X$ and $Y$ discrete, any set of a discrete space would be open, so we would still be allowed to consider the continuous functions $f$ and $h$, to support the homotopic equivalence from $X$ to $Y$.
\\
In the previous subsection we showed that it is possible to obtain the effective curvature of a portion $X$ of the D-brane.
In fact, as mentioned, the parallelism defect mimics the Riemann tensor, from which it is theoretically possible to obtain the corresponding metric.
\\
Therefore, considering the dynamics of the D-brane (gluing and degluing of PNDP-manifolds), and assuming its continuous deformation (described above) in the $Y$ portion, it is then possible to consider the relative continuous deformation of the closed loop and therefore construct an effective curvature (with effective metric) also for the $Y$ portion. The continuous deformation of the closed loop, we consider it as a homotopy between the two closed curves, i.e., we consider $i: \mathbb {R} \rightarrow \mathbb {R}^n$, the closed curve on $X$ and $l: \mathbb {R} \rightarrow \mathbb {R}^n$ the closed curve on $Y$. The homotopy is $s: \mathbb {R} \times [0, 1] \rightarrow \mathbb {R}^n$, such that $s(u, 0) = i(u)$ and $s(u, 1) = l(u)$ we consider send the closed loop $i$ of $X$, with continuity in closed loop $l$ on $Y$. 
\\
At this point, the passage from the closed loop $i$ on $X$ to the closed loop $l$ on $Y$, involves a possible change in the parallelism defect. That is, from the metric of the $X$ portion (which we call $^{X}g_{jk}$), let's move on to the metric of $Y$ (which we call $^{Y}g_{jk}$), and similarly, we can consider this step as a continuous deformation of one into the other, that is, in which each component of the effective metric tensor of the portion $X$ is a function that continuously deforms into the corresponding component of the effective metric tensor of the $Y$ portion.
We obtain an effective metric tensor $g(\xi)_{jk}$, whose components depend on the function $\xi:(u, t) \rightarrow [0, 1]$ (where $u$ is the spatial variables $u_1, u_2, .., u_D$ and $t$ is the time), i.e., for $\xi = 0$ it will correspond to the effective metric tensor of the D-brane portion $X$ (then $g(0)_{jk}=^Xg_ {jk}$), while for $\xi = 1$ it will correspond to the effective metric tensor of the D-brane portion $Y$ (then $g(1)_{jk}=^Yg_ {jk}$). 
\\
So our action, for a fixed dimension, will be of form: 
\\
\begin{equation}
\mathcal{S} \propto \int \mathcal{E}(\ddot{\xi}, N, N^{i}) N{\sqrt{-g(\xi)}d^DVdt}, \label{eq:ouraction}
\end{equation}
\\
where $d^DV=du_1 du_2 du_3...du_D$, $-g(\xi)=-det(g(\xi)_{jk}$), and $\mathcal{E}(\ddot{\xi}, N, N^{i})$ is the proper energy density (energy per unit volume) of the glued PNDPs (analogous to a binding energy) where with $\ddot{\xi}$ we mean a function of  $\xi$  and its derivatives. The ``deformation function", $\xi$, can depend on the brane coordinates, $u_{i}$ and the time, $t$. The quantities $N$ and $N^{i}$ are the ADM lapse and shift assuming that the brane metric, when also considering time evolution, may be written in ADM form. 
\\
\\
Therefore the $S$ action measures the hyper-volume variation due to the dynamics of the D-brane portion (i.e. we consider a transformation of the $X$ portion into the $Y$ portion) obtaining a variation of the extension of its hyper-volume.
\\
\\
We emphasize here that the action (\ref{eq:ouraction}) is an effective action (at the macroscopic level) that governs the dynamics of brane formation which, at the microscopic level, is due to the bonding and detachment of the PNDPs that form and alter the brane. It is not necessarily the full effective gravitational action. Recall that gravitation is associated with the defect of parallelism (the effective holonomy), and summarized again below.
\\
\\
As an example, we can consider a 2-brane (we specify that the dimension chosen in this example is only for computational simplicity, but we could consider D-brane of any dimension), and assume that following the defect in parallelism, and therefore its effective Riemann tensor and effective metric tensor, the continuous deformation from the $X$ portion to the $Y$ portion of the 2-brane (therefore also the deformation of the closed loop), corresponds to the following continuous deformation of the respective components of the effective metric tensors: 
\\
\\
$g(\xi)_{jk}=\left[\begin{matrix}(1-\xi)({^X}g_{11})+\xi ({^Y}g_{11})& (1-\xi)({^X}g_{12})+\xi ({^Y}g_{12})\\ (1-\xi)({^X}g_{21})+\xi ({^Y}g_{21})& (1-\xi)({^X}g_{22})+\xi ({^Y}g_{22})\end{matrix}\right]$.
\\
\\
So the brane action will be:
\\
\begin{equation}
\Scale[0.92]{ \mathcal{S}\propto \int \mathcal{E}(\ddot{\xi},N, N^{i}) N {\sqrt{[(1-\xi)({^X}g_{12})+\xi ({^Y}g_{12})][(1-\xi)({^X}g_{21})+\xi ({^Y}g_{21})]-[(1-\xi)({^X}g_{11})+\xi ({^Y}g_{11})][(1-\xi)({^X}g_{22})+\xi ({^Y}g_{22})}]d^2Vdt}.}
\end{equation}
\\
We obtain that the action is proportional to a variation in area, that is, between the area of the $X$ portion and the area of the $Y$ portion. 
\\
At this point we can summarize in the following points:
\\
\\
(I) We choose a closed loop and, considering it in a portion of the D-brane, we determine a parallelism defect, as described in the previous subsection. With ``portion of D-brane" we mean an open subset of the D-brane, as if it were a manifold in its own right.
\\
(II) We obtain the effective Riemann tensor and derive the corresponding effective metric tensor, assuming, for example, the Levi-Civita connection as the effective connection. The effective metric's components change as a function of $\xi$, 
\\
\\
(III) We write the brane action.
\\
\\
General covariance of the brane action under coordinate transformations which also involve the bulk coordinates off of the brane in this paradigm is an open issue at this stage, as the action (\ref{eq:ouraction}) is intrinsic to the brane and its time evolution.
\\
\\
This gives rise to a type of topological theory governing the formation of branes and that can be related to topological and affine structures of gravity. See \cite{Bajardi, Lombardo1}
\\

\section{ Final Remarks and Conclusions}

A PNDP-manifold considers the concept of virtual dimensions and their resulting observable properties.
In fact, having chosen a manifold and defining a certain algebra on it, the manifold in respect to that algebra will admit a dimension that we call virtual. This is because if we don't look at that algebra and instead consider the underlying manifold, the value of its dimension will be different (see \cite{pndp}). So, in a PNDP-manifold, depending on what is considered, one gets different dimensional results. In fact, this approach suggests that nature manifests itself fundamentally not necessarily as we observe it. This is somewhat like the situation in gauge theories where internal symmetries with gauge freedom give rise only to observables which are gauge invariant. In this respect, PNDP-manifold could be used as a tool to reveal the emergent dimensional aspect of nature, i.e., the emergent space.
\\
In this approach we derived  that in this interpretation, the 3-dimensional manifold, has a virtual negative dimension which interacts with one of the other dimensions such that the manifold emerges as a 1-dimensional vibrating manifold topologically equivalent to a string attached to a point-like manifold. In fact years ago even Arkani-Hamed hypothesized that the space is not a basic property of the dimensions. He said that the space is instead a secondary property created by other more fundamental forces, and in this sense, the dimensions could also vanish, because non-gravitational extra dimensions can be generated dynamically from fundamentally four-dimensional gauge theories (see for example \cite{Arkani-Hamed}, \cite{Hill} and \cite{Cheng}). Therefore, our negative virtual dimension could possibly correspond to a mathematical description of a dimension in which one type of particular fundamental forces allow other dimensions to emerge \cite{Lombardo1,Lombardo2, Myers}.
\\
The ``classical" string theory foresees that originally we have vibrating strings \cite{Witten,Green}. Here instead, in the emerging approach, assume that we originally have these special PNDP-manifolds $I_1 \times I_2 \times (I_3 + E)$ (in case of the open strings) and $S^1_1 \times S^1_2 \times (I_3 + E)$ (in case of the closed strings). These are the manifestation of a fundamental force that generates a structure that has a manifold with a dimension we call virtual negative, and which we describe mathematically with $(I_3 + E)$. This interaction, within the structure, makes a one-dimensional manifold emerge topologically equivalent to an open or closed string \cite{Schwarz}. In the present work we have explored the formation of D-branes and the emergence of gravity on the D-branes via the point-like submanifold of the PNDPs.  In fact, we assert that gravity depends on what we define to be a ``topological defect'' of space. That is, we advance the hypothesis that gravity is a feature inherent in the fabric of the D-brane, and that it, combined with the underlying interactions of the point-like manifolds, can give the illusion of an effective curvature. In other words, we try to consider a scenario in which curved space is  an effect that a flat D-brane can mimic, highlighting a new possible link between gravity and curvature, which relates topological defects (found at a microscopic level) with parallelism defects (found at macroscopic level). Finally, we proceed by developing an action that  describes the internal dynamics of the D-brane and therefore the gluing and ungluing process that takes place in it; a process that we consider by the presence of a variation of the energetics of the brane's hyper-volume.
\\
\\
The role of the string dimension and its possible interactions with the branes, possibly representative of fundamental matter or gauge fields, as in standard string theory, may be considered in future.  From an observational point of view, this scenario could be connected to the so called swampland conjecture giving rise to possible observational constraints \cite{swampland, Benetti}.
\\
\\
\\
\\
\\
\\
\\
\\
{\bfseries \centerline{Acknowledgments}}

\textit{The work was partially funded by Slovak Grant Agency for Science VEGA under the grant number VEGA 2/0009/19.}


\end{document}